\documentclass[]{pasj01}

\usepackage{natbib}

\Received{}
\Accepted{}
 
\begin{document} 

\title{Subaru Hyper Suprime-Cam View of Quasar Host Galaxies at $z<1$ }


\author{Toru \textsc{Ishino}\altaffilmark{1}$^{*}$}
\email{ishino@cosmos.phys.sci.ehime-u.ac.jp}

\author{Yoshiki \textsc{Matsuoka}\altaffilmark{2}$^{*}$}
\email{yk.matsuoka@cosmos.ehime-u.ac.jp}

\author{Shuhei \textsc{Koyama}\altaffilmark{2}}
\author{Yuya \textsc{Saeda}\altaffilmark{1}}
\author{Michael A. \textsc{Strauss}\altaffilmark{3}}
\author{Andy D. \textsc{Goulding}\altaffilmark{3}}
\author{Masatoshi \textsc{Imanishi}\altaffilmark{4,5}}
\author{Toshihiro \textsc{Kawaguchi}\altaffilmark{6}}
\author{Takeo \textsc{Minezaki}\altaffilmark{7}}
\author{Tohru \textsc{Nagao}\altaffilmark{2}}
\author{Akatoki \textsc{Noboriguchi}\altaffilmark{1}}
\author{Malte \textsc{Schramm}\altaffilmark{4}}
\author{John D. \textsc{Silverman}\altaffilmark{8}}
\author{Yoshiaki \textsc{Taniguchi}\altaffilmark{9}}
\author{Yoshiki \textsc{Toba}\altaffilmark{10,11,2}}

\altaffiltext{1}{Graduate School of Science and Engineering, Ehime University, Bunkyo-cho, Matsuyama, Ehime 790-8577, Japan} 
\altaffiltext{2}{Research Center for Space and Cosmic Evolution, Ehime University, 2-5 Bunkyo-cho, Matsuyama, Ehime 790-8577, Japan}
\altaffiltext{3}{Department of Astrophysical Sciences, Princeton University, Princeton, NJ 08540, USA}
\altaffiltext{4}{National Astronomical Observatory of Japan, Mitaka, Tokyo 181-8588, Japan}
\altaffiltext{5}{Department of Astronomical Science, Graduate University for Advanced Studies (SOKENDAI), Mitaka, Tokyo 181-8588, Japan}
\altaffiltext{6}{Department of Economics, Management and Information Science, Onomichi City University, Onomichi, Hiroshima 722-8506, Japan}
\altaffiltext{7}{Institute of Astronomy, The University of Tokyo, Mitaka, Tokyo 181-0015, Japan}
\altaffiltext{8}{Kavli Institute for the Physics and Mathematics of the Universe, WPI, The University of Tokyo, Kashiwa, Chiba 277-8583, Japan}
\altaffiltext{9}{The Open University of Japan, Wakaba 2-11, Mihama-ku, Chiba 261-8586, Japan}
\altaffiltext{10}{Department of Astronomy, Kyoto University, Kitashirakawa-Oiwake-cho, Sakyo-ku, Kyoto 606-8502, Japan}
\altaffiltext{11}{Academia Sinica Institute of Astronomy and Astrophysics, 11F of Astronomy-Mathematics Building, AS/NTU, No.1, Section 4, Roosevelt Road, Taipei 10617, Taiwan}


\KeyWords{galaxies: active --- galaxies: evolution --- galaxies: nuclei --- quasars: general --- quasars: supermassive black holes}  

\maketitle

\begin{abstract}
Active galactic nuclei (AGNs) are key
for understanding the coevolution of galaxies and supermassive black holes (SMBHs). 
AGN activity is thought to affect the properties of their host galaxies, 
via a process called ``AGN feedback", which drives the co-evolution. 
From a parent sample of 1151 $z<1$ type-1 quasars from 
the Sloan Digital Sky Survey quasar catalog,
we detected host galaxies of 862 of them 
in the high-quality $grizy$ images of the Subaru Hyper Suprime-Cam (HSC) survey.
The unprecedented combination of the survey area and depth allows us to 
perform a statistical analysis of the quasar host galaxies, with small sample variance.
We fit the radial  image profile of each quasar
as a linear combination of the point spread function and the $\mathrm{S\acute{e}rsic}$ function,
decomposing the images into the quasar nucleus and the host galaxy components. 
We found that the host galaxies are massive, with stellar mass $M_{\mathrm{star}} \gtrsim 10^{10}M_{\odot}$,
and are mainly located on the green valley. 
This trend is consistent with a scenario in which star formation of the host galaxies is suppressed by AGN feedback, 
that is, AGN activity may be responsible for the transition of these galaxies from the blue cloud to the red sequence.
We also investigated the SMBH mass to stellar mass relation of the $z<1$ quasars,
and found a consistent slope with the local relation, while the SMBHs may be slightly undermassive.
However, the above results are subject to our sample selection, which biases against host galaxies with low masses and/or
large quasar-to-host flux ratios. 
\end{abstract}

\section{Introduction}

Active galactic nuclei (AGNs) and quasars\footnote{
Quasars are the brightest population of AGNs, classically defined with $M_{B} < -23$ mag \citep{SG83, Zakamska03}.}
radiate enormous amounts of energy 
by mass accretion onto supermassive black holes (SMBHs) at the centers of galaxies. 
Previous studies found that 
the mass of SMBHs ($M_{\mathrm{BH}}$) correlates tightly with the stellar velocity dispersion 
or the bulge mass ($M_{\mathrm{bulge}}$) of their host galaxies \citep{Magorrian98, MF01, MD02, HR04, KH13}. 
It was also found that the black hole accretion rate density has a similar shape of redshift evolution 
with the star formation rate (SFR) density, through cosmic time
\citep[e.g.,][]{Silverman08b, Aird10, Aird15, MD14}.
These correlations suggest that SMBHs and their host galaxies may co-evolve. 
Feedback from the AGN has been proposed 
to be responsible for this coevolution \citep[e.g.,][]{Fabian12}.
Simulations suggest that 
the energy released from AGNs could suppress 
star formation activity within their host galaxies, 
by expelling cold gas from galaxies 
via winds driven by radiation pressure  
\citep[``quasar mode'';][]{DiMatteo05}
and/or heating the circum-galactic medium 
\citep[``radio mode'';][]{Croton06}.
For example, hydrodynamical simulations by \citet{Springel05} found that AGN feedback
quenches star formation activity and rapidly produces red ellipticals,
naturally leading to the bimodal distribution of galaxies in the color-magnitude diagram (CMD),
i.e., the blue cloud and the red sequence.
Based on the $\tt{EAGLE}$
\citep[Evolution and Assembly of GaLaxies and their Environments;][]{Schaye15}
hydrodynamical simulations, \citet{Scholtz18} found that 
the observed broad width of specific SFR (sSFR) distribution of galaxies
cannot be reproduced without the effect of AGN feedback.
It is therefore believed that AGNs represent an important phase of galaxy evolution. 
On the observations side, powerful outflows from AGNs are often seen
\citep{Greene12, Liu13a, Liu13b, Cicone14, Brusa18}, which may support 
the idea of quenching of star formation via AGN feedback.
But there is also counter-evidence that 
such outflows do not affect the cold gas reservoir of the host galaxies \citep{Toba17}. 
In addition, other mechanisms such as
common supply of cold gas and stellar/supernova feedback
\citep[e.g.,][]{Angles-Alcazar17a, Angles-Alcazar17b},
and their combinations,
are suggested to explain the observed correlations between SMBHs and their host galaxies.
Therefore, the actual impact of AGNs on their host galaxies is still unclear. 

Direct investigation of AGN host galaxies is one of the most direct ways 
to reveal the impacts of the AGNs on their host galaxies. 
The hosts of type-2 AGNs are relatively easy to study, 
because the strong nuclear radiation at optical wavelengths 
is obscured by the dust torus.
In the local universe, type-2 AGNs in the Sloan Digital Sky Survey 
\citep[SDSS;][]{York00},
selected by the BPT diagram \citep{Baldwin81}, have been studied extensively. 
\citet{Kauffmann03} found that low-luminosity AGNs at $z<0.3$ are typically hosted
in red elliptical galaxies,
while high-luminosity AGNs are hosted in blue elliptical galaxies. 
\citet{Salim07} also found that 
the hosts of more luminous AGNs have higher SFRs. 
On the other hand, \citet{Schawinski10} found that 
AGN host galaxies are located in the green valley, 
i.e., between the blue cloud and the red sequence. 
This result suggests that the host galaxies might have their star formation
being quenched due to AGN feedback.
However, \citet{Jones16} pointed out that 
the BPT selection misses low to moderate luminosity AGNs in star-forming galaxies,
due to dilution of AGN signatures in the emission lines,
so the above results may be biased towards strong AGNs.
\cite{Trump15} also pointed out the same selection bias,
and found that the AGN accretion rates actually correlate with sSFR of their hosts, when the 
selection bias is corrected for.
Their results may indicate that
AGN and star formation activities are triggered 
by the same cold gas reservoir.

At higher redshifts ($z\sim1$), 
investigation of the hosts of X-ray detected obscured AGNs
\citep{Nandra07, Georgakakis08, Silverman08a}
suggest that the host galaxies are located in the red sequence and the green valley. 
On the other hand, \citet{Cardamone10} found that 
most of the host galaxies of X-ray AGNs at $z<1$ in the green valley 
are intrinsically blue galaxies reddened by dust. 
It has also been suggested that 
the AGN fraction in a luminosity-limited sample becomes
high around the green valley or the red sequence, 
while those in a stellar mass-limited sample becomes 
high around the blue cloud or is uniform across colors \citep{Silverman09, Xue10}.

\citet{Hickox09} found that the hosts of AGNs selected in infrared (IR), X-ray, or radio 
are located in different regions of the CMD. 
This may indicate that different types of AGNs represent different evolutionary stages of galaxies. 
It is therefore important to investigate the hosts of various types of AGN.

This study focuses on optically selected luminous type-1 quasars. 
In contrast to type-2 objects, the host galaxies of type-1 quasars are difficult to observe, 
because their radiation is outshone by the bright central nuclei \citep[e.g.,][]{Matsuoka14}. 
It is necessary to accurately decompose the light of quasars into central nuclei and host galaxies. 
We adopt an image (surface brightness profile) decomposition method, 
using the point spread function (PSF) to model the nuclear radiation.
While high-spatial-resolution data from the $\it{Hubble\ Space\ Telescope}$ are most useful 
for such decomposition \citep[e.g.,][]{Bahcall97, Jahnke04, Sanchez04, Villforth17}, 
the sample sizes are usually limited. 
In order to perform a statistical analysis, high-quality images of a large number of quasars are necessary. 
\citet{Matsuoka14} investigated the stellar population of 
about 800 SDSS type-1 quasar hosts at $z<0.6$, 
using co-added images on SDSS Stripe 82 \citep{Annis14}, 
which are $\sim2$ mag deeper than the normal SDSS images.
They found that quasar host galaxies are located on the blue cloud
to the green valley, and are absent on the red sequence.

This study aims to extend the analysis of \citet{Matsuoka14} to higher redshift and lower host luminosities,
by exploiting the  data taken by the Subaru Hyper Suprime-Cam \citep[HSC;][]{Miyazaki18}.
Deep imaging with superior seeing offered by HSC
is suitable for studies of quasar host galaxies;
the bright nuclear radiation is confined to
a small number of CCD pixels,
and detection of the surrounding faint galaxies
benefits from low noise level.
We perform image decomposition of SDSS quasars 
and extract the host galaxy components, 
and analyze their properties. 
This paper is organized as follows. 
In Section 2, we describe our data and sample selection. 
We describe the image decomposition 
and the spectral energy distribution (SED) fitting in Section 3. 
The results and discussion appear in Section 4. 
The summary is presented in Section 5. 
Throughout this paper, we use the cosmological parameters of 
$H_{0}=70\mathrm{\ km\ s^{-1}\ Mpc^{-1}},\ \Omega_{\mathrm{M}}=0.3$, 
and $\Omega_{\Lambda}=0.7$. 
All magnitudes are presented in the AB system \citep{Oke1983} 
and are corrected for Galactic extinction \citep{Schlegel98}.

\section{Data and Sample}

\subsection{HSC Imaging Data}

This study uses imaging data and the source catalog obtained 
through the HSC Subaru Strategic Program (HSC-SSP) survey \citep{Aihara18}. 
The HSC is a wide-field camera installed on the Subaru 8.2 m telescope \citep{Miyazaki18}. 
It is equipped with 116 2K $\times$ 4K CCDs, 
and has a pixel scale of $0\farcs168$ 
and a field of view of 1.5 deg in diameter. 
The HSC-SSP survey has three layers (Wide, Deep, and UltraDeep). 
This study uses Wide layer data covering about 300 deg$^{2}$,
included in the S17A internal data release \citep{Aihara19}. 
The Wide layer is observed with five broad bands 
($g, r, i, z, \mathrm{and}\ y$). 
The $5\sigma$ limiting magnitudes measured within $2^{\prime\prime}$ aperture are 
$g=26.5$, $r=26.1$, $i=25.9$, $z=25.1$, and $y=24.4$ mag.
The typical seeing values are $0\farcs72$, $0\farcs67$, $0\farcs56$, $0\farcs63$, $0\farcs64$
in the $g,\ r,\ i,\ z,\ {\rm{and}}\ y$ bands, respectively. 
For reference, SDSS Stripe 82 \citep[used by][]{Matsuoka14} has 
a 5$\sigma$ limiting magnitude of $r \sim 24$ mag and seeing $\sim 1\farcs1$. 

\begin{figure}
\begin{center}
\includegraphics[width=9cm]{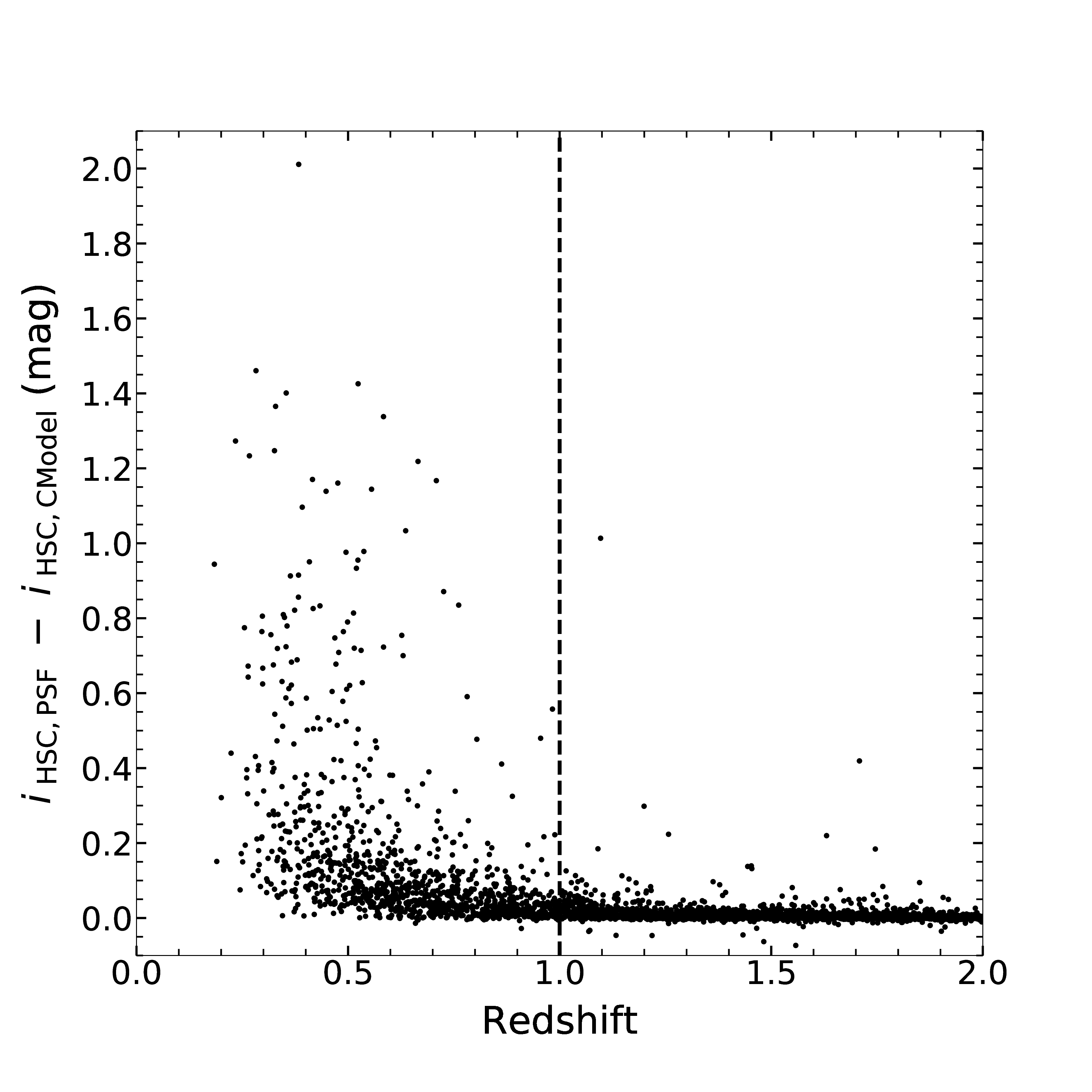} 
\end{center}
\caption{
Difference between the HSC PSF magnitudes and CModel magnitudes of the matched HSC-SDSS quasars in the $i$-band, 
as a function of redshift. The dashed line represents $z=1$, the maximum redshift of the present sample.
}
\label{extend}
\end{figure}

\begin{figure*}
\begin{center}
\includegraphics[width=9cm, angle=-90]{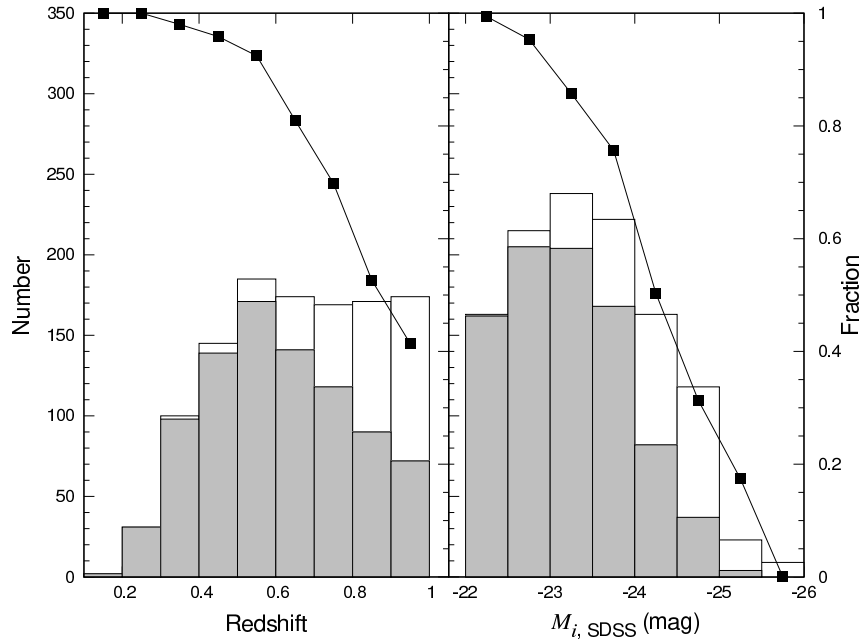} 
\end{center}
\caption{
The white and gray histograms represent the distributions of the matched HSC-SDSS quasars at $z<1$
and our main sample quasars, respectively, 
as a function of redshift (left) and $M_{i,\mathrm{SDSS}}$ (right). 
The ratios of the number of the main sample quasars to that of the matched HSC-SDSS quasars in each bin are represented 
by the squares connected by the solid lines (the values are shown in the right axis).
}
\label{num_dist}
\end{figure*}

\subsection{Sample selection \label{subsec:sample_selection}}
We used the SDSS data release 7 \citep[DR7;][]{Abazajian09} quasar catalog \citep{Schneider10} 
to select objects known spectroscopically to be quasars. 
This quasar catalog covers about 9380 deg$^{2}$ and 
contains 105783 quasars with $M_{i,\ \mathrm{SDSS}}$
\footnote{$M_{i,{\mathrm{SDSS}}}$ is an absolute magnitude based on a SDSS PSF magnitude, 
$k$-corrected to $z=0$ \citep[][]{Schneider10}.}$<-22$ mag. 
We searched in the HSC catalog database for a counterpart to each SDSS quasar, within $1^{\prime\prime}$.
In the HSC catalog query, we excluded sources affected by cosmic rays or nearby bright objects,
by setting $\tt{pixelflags.crcenter=False}$ and $\tt{pixelflags.bright.objectcenter=False}$.
We also excluded sources which are saturated ($\tt{pixelflags.saturatedcenter=False}$),
close to the edge of the CCD ($\tt{pixelflags.edge=False}$), detected on bad pixels ($\tt{pixelflags.bad=False}$),
blended ($\tt{nchild=0}$), or duplicates ($\tt{isprimary=True}$).
This resulted in 4684 SDSS quasars at $z \lesssim 5$ matched to HSC sources.
Seven of the quasars were matched to two HSC sources within $1^{\prime\prime}$,
and we selected the nearest HSC source as a counterpart in each of these cases.

We first checked the extendedness of the matched quasars
and determined the redshift range of the present analysis,
in which a significant fraction of the quasars have detectable host galaxies. 
Figure \ref{extend} shows the  difference between the PSF
and CModel magnitudes ($i_{\mathrm{HSC,PSF}} - i_{\mathrm{HSC,CModel}}$)
of the HSC-SDSS matched quasars
in the HSC $i$-band, as a function of redshift. 
The PSF magnitudes are measured by fitting the PSF,
and the CModel magnitudes are measured by fitting two-component, 
two-dimensional galaxy profiles (exponential and de Vaucouleurs profiles) 
to the source image
\citep{Abazajian04, Bosch18}. 
The galaxy model profiles are convolved with the PSF
measured in the HSC imaging precessing pipeline, $\tt{hscPipe}$ \citep{Bosch18}. 
The magnitudes are obtained by integrating the best-fit models to the radius infinity.
Extended sources have brighter CModel magnitudes than PSF magnitudes,
because the former captures the extended radiation correctly with the adopted galaxy model profiles,
while the latter captures only the central radiation within PSF.
Therefore, the difference between the PSF and Cmodel magnitudes 
is a good measure of source extendedness. 
As clearly shown in Figure \ref{extend}, $i_{\mathrm{HSC,PSF}} - i_{\mathrm{HSC,CModel}}$ have relatively larger values at the lowest $z$,
and converge to zero at $z > 1$.
We measured the standard deviation ($\sigma$) of the $i_{\mathrm{HSC,PSF}} - i_{\mathrm{HSC,CModel}}$ values for the quasars at $z > 3$
(which are considered to be point sources) 
and found that $\sim$70 \% of the quasars at $z < 1$ have
$i_{\mathrm{HSC,PSF}} - i_{\mathrm{HSC,CModel}}$ values larger than 3$\sigma$. 
On the other hand, only 4 \% of the quasars at $z > 1$ satisfy this criterion.
Thus we decided to focus on the 1151 SDSS quasars with HSC imaging data at $z < 1$ in this work.

This paper also uses the following quantities
taken from \citet{Shen11}.
The extinction-uncorrected continuum luminosities 
at rest-frame 3000 \AA\ ($L_{3000}$)
and $[\mathrm{OI\hspace{-.1em}I\hspace{-.1em}I}]\ \lambda5007$ line luminosities
($L_{[\mathrm{OI\hspace{-.1em}I\hspace{-.1em}I}]}$) 
are used as tracers of quasar nuclear power \citep[e.g.,][]{Kauffmann03}.
The contribution of the host galaxy to $L_{3000}$ is negligible,
since the nucleus is much brighter 
in the ultraviolet part of a quasar spectrum \citep[e.g.,][]{Selsing16}. 
We also use the fiducial values of $M_{\mathrm{BH}}$ reported in \citet{Shen11},
which were measured from H$\beta$ emission lines for $z < 0.7$ quasars \citep[][]{VP06} 
and from Mg I\hspace{-.1em}I emission lines for $z > 0.7$ quasars. 
The Mg I\hspace{-.1em}I- and H$\beta$-based $M_{\mathrm{BH}}$ are found to be consistent on average;
the mean offset and $1\sigma$ scatter between the two calibrations are 
0.009 dex and 0.25 dex, respectively \citep{Shen11}.
55 \% of our quasars have $M_{\mathrm{BH}}$ from H$\beta$,
and 45 \% have $M_{\mathrm{BH}}$ from Mg I\hspace{-.1em}I.

\section{Analysis}

\subsection{Image Decomposition \label{subsec:image}}

Our analysis uses a similar method to that presented in \citet{Matsuoka14}. 
We used HSC cutout images in the five bands around each quasar. 
From these images, we constructed azimuthally averaged radial profiles 
centered on the quasar coordinates. 
We defined annular bins of one-pixel width out to radius $R=25$ kpc,
and measured the mean flux and its error ($\sigma(R)$) in each bin. 
The error is the standard deviation of pixel counts ($\sigma_{\mathrm{std}}(R)$) 
divided by the square root of the number of pixels included in the bin.
To remove contamination by nearby objects,
cosmic ray signals, and so on,
we performed 3$\sigma_{\mathrm{std}}(R)$ clipping
when calculating the mean fluxes.
The observed flux level is indistinguishable from the sky background at $R > 25$ kpc,
hence the present analysis encompasses all the detectable radiation from the individual quasars.	
Although sky subtraction has already been performed
by $\tt{hscPipe}$ \citep{Bosch18}, 
we found that there are small sky subtraction residuals in the HSC images. 
Therefore, we subtracted the mean value of pixel counts in the annulus 
between $25-30$ kpc from the quasar as a residual sky.

We fitted the measured radial profile 
with a linear combination of PSF 
and a \citet{Sersic} function 
to decompose it into nuclear and host galaxy components. 
We utilized the PSF models created by $\tt{hscPipe}$, 
which uses a modified version of the $\tt{PSFEx}$ code \citep{Bertin13}.
On average, 70 stars were selected on each CCD to model the PSF.  
The light distributions of these stars on the images are fitted as a function of position, 
so that PSFs can be predicted at any position on each CCD.
The $\mathrm{S\acute{e}rsic\ function}$ has the following form
	\begin{eqnarray}
	\label{Sersic}
		I_{\mathrm{S\acute{e}rsic}}(R) &=& I_{e} \exp \Big(-b_{n} \Big[ \Big( \frac{R}{R_{e}} \Big)^{1/n} -1 \Big ] \Big), 
	\end{eqnarray}
where $R_{e}$ is the effective radius including half of the total flux, and $I_{e}$ is the intensity at the radius $R=R_{e}$. 
The $\mathrm{S\acute{e}rsic}$ index $n$ determines the shape of the profile
and $b_{n}$ is a constant determined for a given $n$
\citep[the relation between $n$ and $b_{n}$ is presented in][]{GD05}.
The exponential and de Vaucouleurs profiles are represented by $n=1$ and $4$, respectively.
The $\mathrm{S\acute{e}rsic\ function}$ is convolved with the PSF. 
We determined the best-fit $\mathrm{S\acute{e}rsic}$ model by minimizing the $\chi^{2}$ value expressed by
	\begin{eqnarray}
		\chi^{2} &=& \Sigma \Big[ \frac{I(R)-(A \times I_{\mathrm{PSF}}(R)+I_{\mathrm{S\acute{e}rsic}}(R))} {\sigma(R)} \Big]^{2},
	\end{eqnarray}
where $I(R)$ is the measured quasar radial profile and the constant $A$ scales the PSF intensity, $I_{\mathrm{PSF}}(R)$. 
We used those radial bins with $I(R) > 3\sigma(R)$ for the profile fitting.

Before performing the actual profile fitting, 
we checked if there was a significant contribution 
from the host galaxy radiation in each quasar, as follows. 
In the $i$-band, we fitted only the PSF model 
to the quasar profile at $R < 2$ pixels 
(corresponding to a diameter of $\sim 0\farcs6$,
comparable to the seeing disk), 
and subtracted this best-fit PSF from the quasar profile.
We removed those quasars dominated by the central PSF, 
with the residual flux less than $10\%$ of the subtracted PSF flux. 
This criterion drops 289 objects, 
and the remaining 862 objects constitute our main sample used throughout the following analysis.
We test performed the following image decomposition 
for the dropped 289 objects,
and found that the resultant host galaxy fluxes are very uncertain
(with the typical error of $\sim1$ mag)
and are therefore not usable. 
Figure \ref{num_dist} shows the distributions of the redshift
and $M_{i,\mathrm{SDSS}}$ of the 1151 HSC-SDSS matched quasars at $z<1$
and those of the finally-selected 862 objects.
The figure also shows the ratios of the number of the main sample quasars
to that of the $z < 1$ HSC-SDSS matched quasars in each bin of redshift and $M_{i,\mathrm{SDSS}}$.
As expected, relatively distant objects ($z\gtrsim0.8$) tend to be dropped 
by the above criterion. 
The brighter ($M_{i,\ \mathrm{SDSS}}<-24$ mag) quasars also tend to be dropped,
as they are more likely to outshine the host galaxies.
Hence the present sample does not contain the most luminous quasars, whose feedback effect on the host galaxies may be most significant.

\begin{figure*}
\begin{center}
\includegraphics[width=12cm, height=20cm]{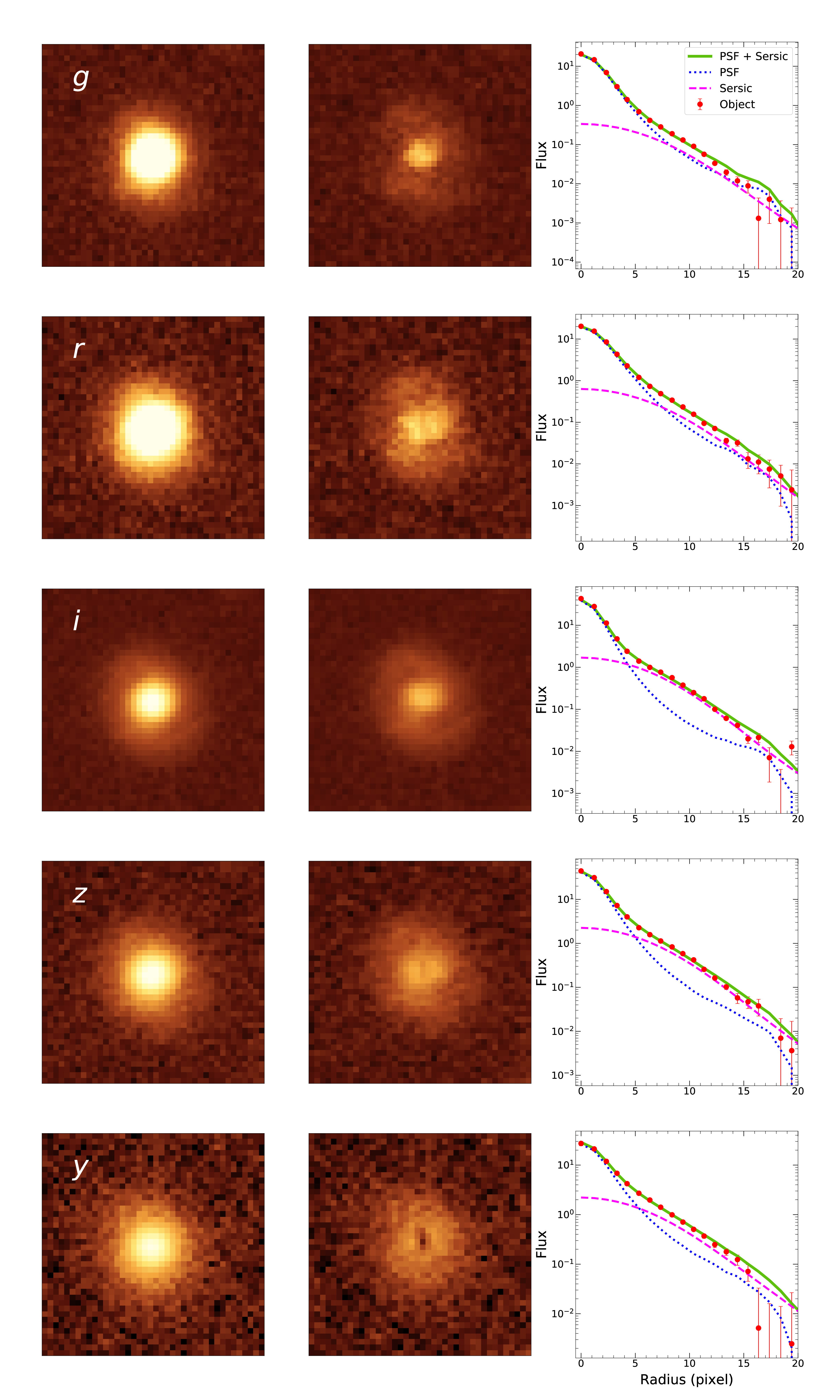} 
\end{center}
\caption{
An example of the decomposition results for one of the quasars, with $M_{i,\ \mathrm{SDSS}}=-23.65$ mag. 
This quasar is at a relatively high redshift (z = 0.824) and provides reasonable fitting results,
giving clear detection of the host galaxy in all five bands.
Left panels show the original $7^{\prime\prime} \times 7^{\prime\prime}$ quasar images. 
Middle panels show the images after subtracting the best-fit PSF components. 
Right panels show the profile fitting results. 
The circles represent the azimuthally averaged radial profile of the quasar. 
The dashed and dotted lines represent the PSF components 
and the $\mathrm{S\acute{e}rsic}$ components, respectively. 
The solid lines represent the best-fit models, 
the sum of the PSF and the $\mathrm{S\acute{e}rsic}$ components.
}
\label{decomposition}
\end{figure*}

Our decomposition method has four free parameters,
i.e., the PSF scaling parameter ($A$) and the three $\mathrm{S\acute{e}rsic}$ parameters
($I_{e}, R_{e}$, and $n$ in Equation (\ref{Sersic})).
In order to avoid parameter degeneracy, 
we divided the profile fitting process into two steps.
First, we fitted $\mathrm{S\acute{e}rsic}$ models at $R>2$ pixels
to the above residual profiles after subtracting the best-fit PSF,
and determined $R_{e}$ and $n$ values;
this was performed in the $i$-band.
The galaxy component is most easily visible in the $i$-band, 
because the $i$-band images were obtained 
under the best seeing conditions \citep{Aihara18}. 
Moreover, the quasar-to-galaxy contrast is smaller at longer wavelengths, 
and the $i$-band is deeper than the $z$ and $y$-bands.
We vary $R_{e}$ from 1 to 20 kpc 
(corresponding to 0.9 to 18 pixels, or $0\farcs15$ to $3''$, at the mean redshift of the sample, $z = 0.59$)
with a grid of 1 kpc, and $n$ from 0.5 to 5.0 with a grid of 0.1.
Second, we simultaneously fitted the PSF 
and $\mathrm{S\acute{e}rsic}$ models
to the original quasar profiles,
with $R_{e}$ and $n$ fixed to the above values, in the five bands.
The parameters $A$ and $I_{e}$ were changed freely in each band.  
Figure \ref{decomposition} shows an example of the original HSC images,
the PSF-subtracted images, and the profile fitting results for one of the quasars.
This quasar is at a relatively high redshift (z = 0.824) and provides reasonable fitting results,
giving clear detection of the host galaxy in all five bands.

\begin{figure}	
\begin{center}
\includegraphics[width=7cm, angle=-90]{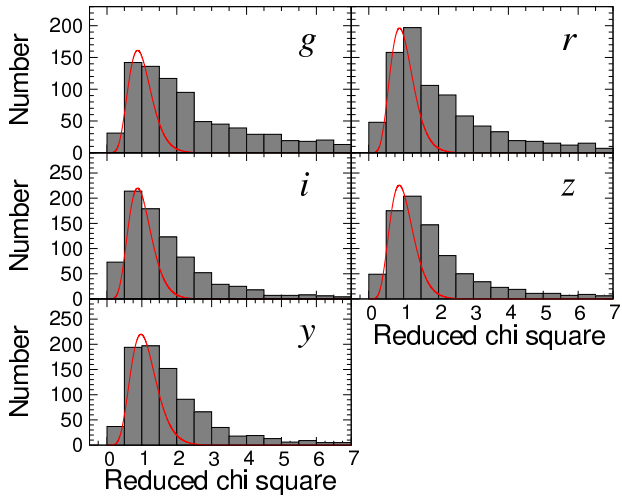} 
\end{center}
\caption{
Distributions of the reduced $\chi^{2}$ values of the radial profile fitting,
for the 862 quasars.
The solid lines represent the expected reduced $\chi^{2}$ distribution,
calculated with the mean degree of freedom in each band
(arbitrarily scaled).
}	
\label{chi}
\end{figure}

\begin{figure*}
\begin{center}
\includegraphics[width=16cm, height=12cm]{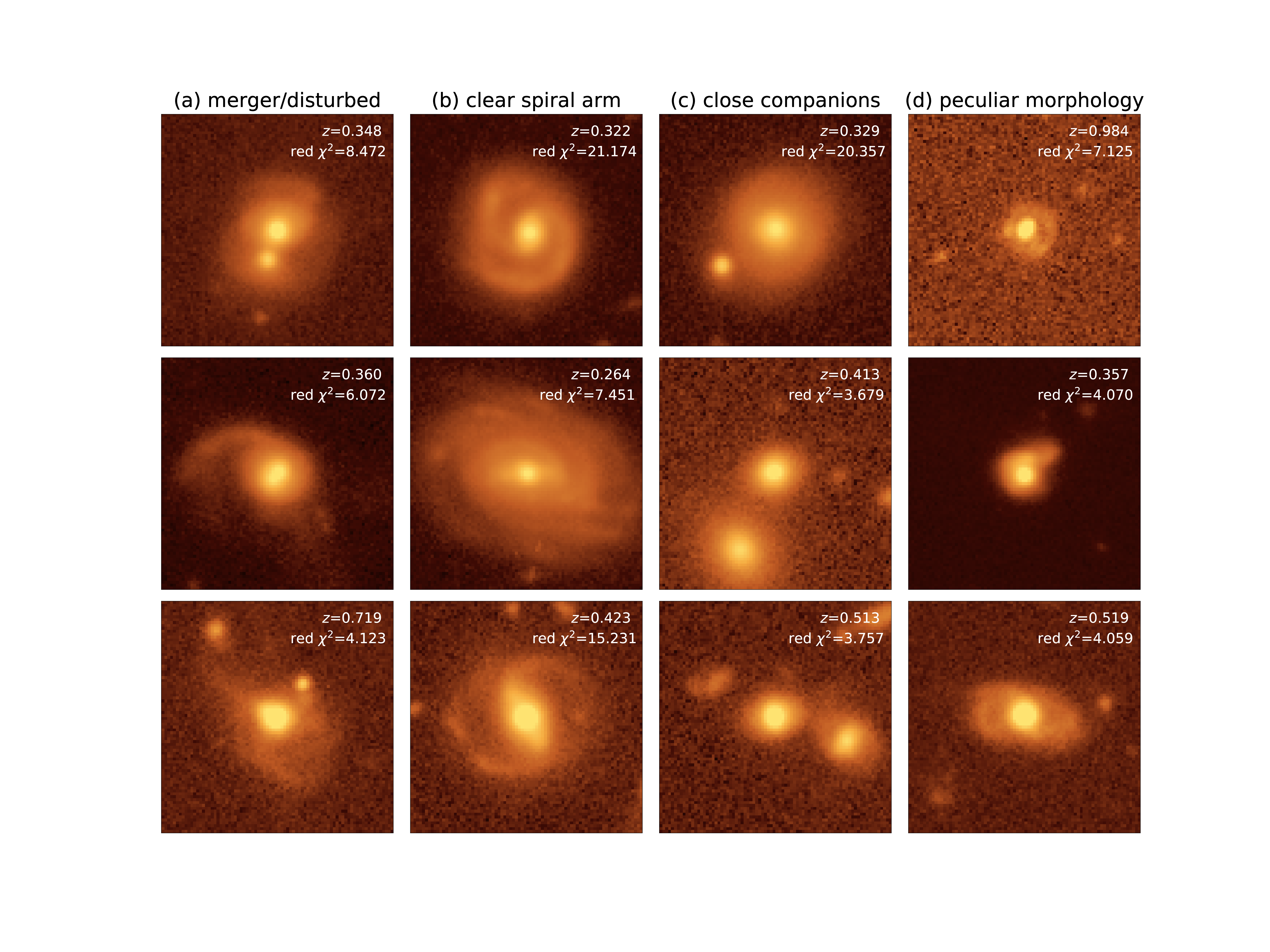} 
\end{center}
\caption{Examples of the $i$-band $13^{\prime\prime} \times 13^{\prime\prime}$ images of the quasars 
with reduced $\chi^{2}\gtrsim3$ 
((a): merger signatures or disturbed morphologies,
(b): clear spiral arms,
(c): close companions,
(d): peculiar morphologies).
The redshift and reduced $\chi^{2}$ in the $i$-band
are shown in top right of each panel.
}
\label{quasar_image}
\end{figure*}

Figure \ref{chi} shows distributions of the reduced ${\chi}^{2}$ values 
of the profile fitting in the five bands. 
In all bands, the distributions have a peak around 1, 
suggesting that our models reproduce the observed profiles reasonably well. 
While they are approximately close to the expected reduced ${\chi}^{2}$ distribution,
the number of objects with reduced $\chi^{2}\gtrsim3$ seems larger than expected.
We checked the $i$-band images of the 138 quasars with reduced $\chi^{2}\gtrsim3$,
and found that 50 \% (69/138) show irregular morphologies
(merger: 9/117, disturbed: 13/117, clear spiral arms: 13/117, 
close companions: 16/117, peculiar morphology: 18/117).
Their example images are shown in Figure \ref{quasar_image}.
Such azimuthally-asymmetric structures cannot be reproduced by our models, 
and would result in large $\chi^{2}$ values.

We calculated the host galaxy flux ($f_{\mathrm{galaxy}}$) 
by integrating the best-fit $\mathrm{S\acute{e}rsic\ profile}$ out to $R=25$ kpc. 
The nuclear flux ($f_{\mathrm{quasar}}$) is also derived 
by integrating the best-fit PSF profile over the same range.

\begin{longtable}{ll}
	\caption{CIGALE parameters}\label{CIGALE}
  	\hline              
  	Parameters & Values \\ 
	\endfirsthead
	\hline
  	Parameters & Values \\
	\endhead
  	\hline 
	\endfoot
  	\hline
	\endlastfoot
  	\hline 
  	Star formation $e$-folding time $\tau$ & 0, 0.1, 0.5, 1, 2, 3, 4, 5, 6, 7, 8, 9, 10, 11, $\infty$ [Gyr] \\
  	Age of the stellar population & 0.05, 0.1, 0.5, 1, 2, 3, 4, 5, 6, 7, 8, 9, 10, 11, 12, 13 [Gyr] \\
	Color excess $E(B-V)$ & 0, 0.1, 0.2, 0.3, 0.4, 0.5, 0.6, 0.7, 0.8, 0.9, 1.0, 1.1, 1.2, 1.3, 1.4 \\
	Metallicity $Z$ & 0.02 (Solar metallicity) \\	
 \end{longtable}

 \begin{figure*}
\begin{center}
\includegraphics[width=10cm, height=20cm]{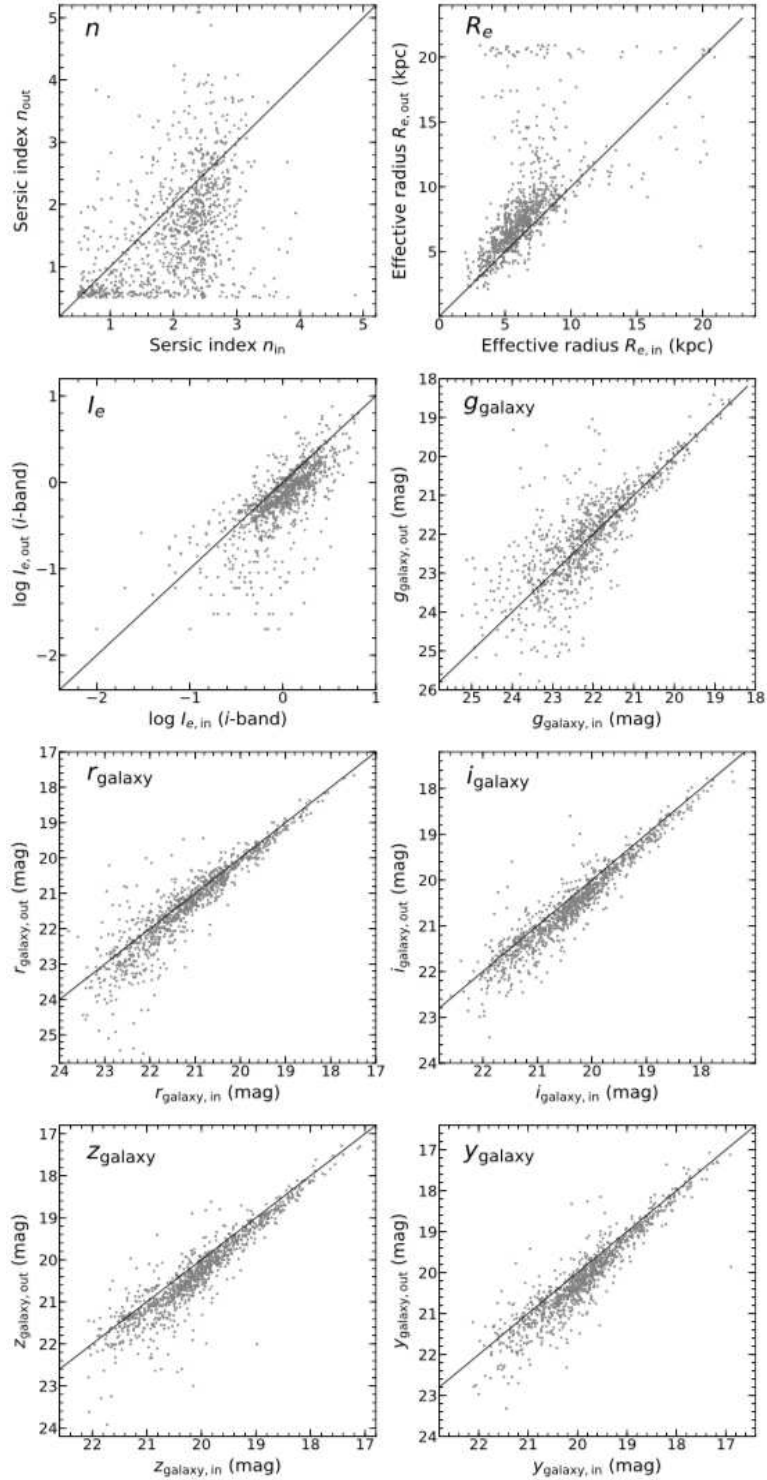} 
\end{center}
\caption{Comparison of the input (derived from galaxy images before superimposing stars to make artificial quasars)
and output values (derived by the decomposition of the artificial quasar images)
of $\mathrm{S\acute{e}rsic}$ index $n$ (top left), 
effective radius $R_{e}$ (top right), the $i$-band $I_{e}$ (second to top left),
and galaxy magnitudes in the $g$ (second to top right), $r$ (third to top left), 
$i$ (third to top right), $z$ (bottom left), and $y$ (bottom right) bands. 
The solid lines represent the locus where the input and output values are equal.
We added small random offsets to each point in the panels of $n$ and $R_{e}$, for visibility.
}
\label{simulation1}
\end{figure*}

\begin{figure*}
\begin{center}
\includegraphics[width=18cm, height=6cm]{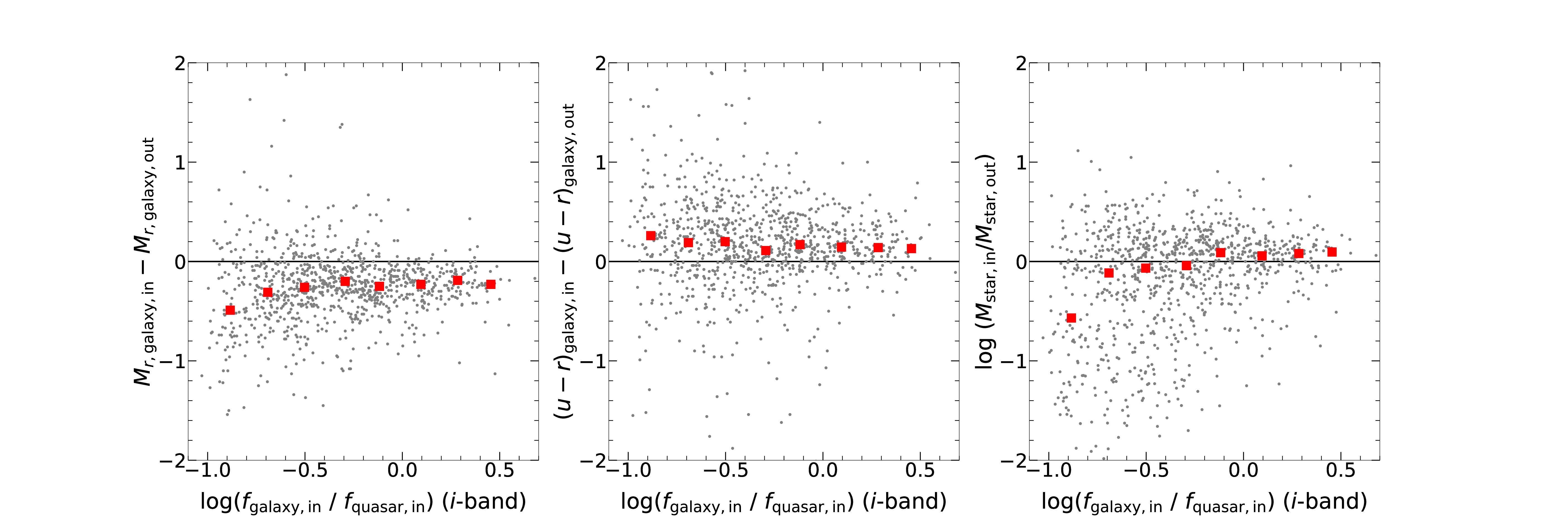} 
\end{center}
\caption{Difference between the input and output values of $M_{r,\mathrm{galaxy}}$ (left),
$(u-r)_{\mathrm{galaxy}}$ (middle), and stellar mass $M_{\mathrm{star}}$ (right),
as a function of the input galaxy-to-quasar flux ratio in the $i$-band,
$f_{\mathrm{galaxy,in}}/f_{\mathrm{quasar,in}}$.
The squares represent the median values 
in (0.2-dex wide) bins of $f_{\mathrm{galaxy,in}}/f_{\mathrm{quasar,in}}$.
The solid lines represent the locus where the input and output values are identical.
}
\label{simulation2}
\end{figure*}

\subsection{SED Fitting}

We performed  SED fitting to the decomposed host galaxy fluxes using $\tt{CIGALE}$ 
\citep[Code Investigating GALaxy Emission;][]{Burgarella05, Noll09, Boquien19}. 
$\tt{CIGALE}$ estimates physical quantities of galaxies 
such as SFR and stellar mass ($M_{\mathrm{star}}$),
by fitting a set of SED models to the observed data. 
We used the same range of SED-fitting parameters as in \citet{Tanaka15}, 
in order to compare the model results with those of non-AGN galaxies 
selected from the HSC data (see Section \ref{subsec:cmd}). 
We created template SEDs using the stellar population synthesis models of \citet{BC03}. 
Stellar ages were varied from 50 Myr to 13 Gyr. 
We assumed an exponentially decaying star formation history (i.e., the $\tau$-model),
which was often used by previous studies \citep[e.g.,][]{Bongiorno12, Illbert13, Muzzin13, Toba18},
and varied $\tau$ from 0.1 Gyr to 11 Gyr.
$\tau=0$ and $\tau=\infty$ models are also constructed to represent instantaneous 
and constant star formation history, respectively. 
The \citet{Calzetti00} dust attenuation curve was applied 
with $E(B-V)$ ranging from 0.0 to 1.4. 
We used the \citet{Chabrier03} initial mass function (IMF)
and assumed solar metallicity.
These fitting parameters are summarized in Table \ref{CIGALE}. 
84 \% of the quasars in our sample have detectable host galaxy components, with
$>3\sigma$ significance in the measured $f_{\mathrm{galaxy}}$,
in more than three bands,
allowing for reasonable SED fitting.

In the following discussion,
we use the $r$-band absolute magnitudes ($M_{r,\mathrm{galaxy}}$)
and the rest-frame SDSS $u$-band minus HSC $r$-band colors 
($(u-r)_{\mathrm{galaxy}}$)
of the host galaxies,
derived by $k$-correcting the decomposed host magnitudes
using the best-fit SED templates.
We also use $M_{\mathrm{star}}$ derived
from the $\tt{pdf \_ analysis}$ module of $\tt{CIGALE}$.
The $r$-band absolute magnitudes of the quasar nuclei 
($M_{r,\mathrm{quasar}}$)
are calculated from the PSF component, 
with the amount of $k$-correction derived from 
the quasar composite spectrum \citep{Selsing16}.

We also test performed SED-fitting with a combination 
of two exponential decaying star formation histories, 
represented by a $\tau$-model plus a late starburst,
but it does not change our results significantly.
Given that we have only five bands for SED fitting,
we adopt the simple one-component $\tau$-model
in the following discussion. 
   
\subsection{Saturated quasars \label{subsec:saturated}}

The SDSS catalog includes quasars which are saturated on HSC images 
and are thus excluded from our main sample.
In order to assess the bias that could be introduced
by excluding such saturated 
(i.e., relatively luminous) quasars,
we analyzed a sample of 127 SDSS quasars
which are flagged as saturated in the HSC database.
The saturated quasars have the comparable typical redshift and $M_{\mathrm{BH}}$
($z=0.61$ and $\log({M_{\mathrm{BH}}/M_{\odot}})=8.57$) to,
but significantly higher typical luminosity ($M_{i,\ \mathrm{SDSS}} = -24.32$ mag) than, the main sample 
($z=0.59$ and $\log({M_{\mathrm{BH}}/M_{\odot}})=8.50$, $M_{i,\ \mathrm{SDSS}} = -23.15$ mag).

In the profile fitting of the saturated quasars, we excluded saturated pixels
identified by the HSC image mask plane.
If all pixels within a given bin is saturated, then we did not use that bin for the profile fitting.
Almost all the $R<2$ pixels are saturated in these quasars,
and so we cannot perform the two-step profile fitting
described in Section \ref{subsec:image}.
Therefore, we simultaneously fitted PSF and $\mathrm{S\acute{e}rsic}$ models 
to the observed radial profiles,
with all the four parameters ($A$, $I_{e}$, $R_{e}$, $n$) varied 
at the same time, in the $i$-band.
Then only $A$ and $I_{e}$ were varied in the other four bands,
with $R_{e}$ and $n$ fixed to those determined in the $i$-band.
The subsequent SED fitting analysis was performed 
in the same way as for the main sample.
The saturated sample was created only to check the bias of excluding those luminous quasars from the present work,
and are thus treated separately from the main sample throughout the remaining analysis.

\subsection{Systematic errors \label{subsec:simulation}}

Before proceeding to the results and discussion,
we evaluate systematic errors that could be induced 
by the methods we described above.
In order to create artificial quasar images,
we extracted spectroscopically-identified (and flux-limited) sample of $z < 1$ galaxies 
and stars from SDSS DR15 \citep{Aguado19}.
For the stars, we applied the same flux range as $f_{\mathrm{quasar}}$ 
derived by the decomposition. 
SDSS classifies the objects into stars, galaxies, and quasars 
based on the spectral template fitting
\footnote{https://www.sdss.org/dr16/spectro/catalogs/}.
Type-2 AGNs could be classified as galaxies,
but this is not a problem for our purpose,
since the nuclear continuum radiation is blocked by the obscured material 
\citep[e.g.,][]{Kauffmann03, Nandra07, Silverman08a}.
For each galaxy, we matched a star in the closest separation;
the matched star is always found within $1^{\prime}$ for $z<0.7$ galaxies
and within $2^{\prime}$ for the entire galaxy sample at $z<1$. 
From the 1970 galaxy-star pairs thus created,
we randomly selected 976 pairs in such a way
that the galaxy redshift distribution becomes the same
as that of the main quasar sample.

The above (spectroscopically confirmed) galaxies tend to be
brighter than the quasar host galaxies we analyzed above.
So we reduced the galaxy brightness by scaling down the image pixel counts,
so that the distribution of the integrated fluxes matches that of the quasar host galaxies.
Since the background noise was also reduced by this procedure,
we added random Gaussian noise
to recover the original noise level.
The brightness of the stars were similarly adjusted,
in such a way that the resultant galaxy-to-star flux ratios
have the same distribution as do
the galaxy-to-nuclear flux ratios of the quasars. 
We superimposed these galaxy and star images by aligning the centers, within $R < 25$ kpc region,
which simulate quasars with the host galaxies.
We note that addition of star images brings in additional background noise, so the simulated quasar images have
a little larger noise than the real quasar images we analyzed.
Finally, we performed the same 1d profile fitting as we did for the real quasar sample.
Hereafter the quantities
derived from the galaxies (i.e., the $\mathrm{S\acute{e}rsic}$ parameters and fluxes) before/after superimposing the stars
are referred to as ``input'/output values'' and are denoted $x_{\mathrm{in}}$/$x_{\mathrm{out}}$.
The input star fluxes are denoted as $f_\mathrm{quasar,in}$.

Figure \ref{simulation1} compares the input 
and output $\mathrm{S\acute{e}rsic}$ parameters
($I_{e},\ R_{e},\ n$, determined in the $i$-band)
and the galaxy fluxes in the five bands. 
While the $\mathrm{S\acute{e}rsic}$ parameters show relatively large scatters,
which is likely caused by the parameter degeneracy in the image decomposition,
the total galaxy fluxes show good correlations.
That is, our decomposition method can extract 
the host galaxy fluxes reasonably well,
even though each $\mathrm{S\acute{e}rsic}$ parameter
is not determined accurately.
The present study does not use the individual $\mathrm{S\acute{e}rsic}$ parameters in any form, 
so their inaccuracy does not impact the following results.
The systematic offsets and standard deviation of the galaxy fluxes in each band
($m_{\mathrm{galaxy,in}}-m_{\mathrm{galaxy,out}}$) are
$0.03\pm0.78$,
$-0.15\pm0.54$,
$-0.19\pm0.31$,
$-0.23\pm0.42$,
and $-0.23\pm0.38$ mag
in the $g$, $r$, $i$, $z$, and $y$-bands, respectively.
Our decomposition method tends to 
underestimate galaxy fluxes by about 0.2 mag,
presumably by oversubtracting
the central galaxy lights with the PSF models. 
Figure \ref{simulation2} shows the offsets of 
$M_{r,\mathrm{galaxy}}$, $(u-r)_{\mathrm{galaxy}}$, and $M_{\mathrm{star}}$,
as a function of the input galaxy-to-quasar flux ratio in the $i$-band
($f_{\mathrm{galaxy,in}}/f_{\mathrm{quasar,in}}$).
Our method estimates $\sim0.2$ mag fainter $M_{r,\mathrm{galaxy}}$
and $\sim0.2$ mag bluer $(u-r)_{\mathrm{galaxy}}$
than the input, almost regardless of the galaxy-to-quasar flux ratio.
The systematic offset of $M_{\mathrm{star}}$ is small, 
except at the smallest $f_{\mathrm{galaxy,in}}/f_{\mathrm{quasar,in}}$ 
(which is likely affected by small number statistics and difficulty in extracting the faintest hosts).
The median of $\log({M_{\mathrm{star,in}}/M_{\mathrm{star,out}}})$ is $-0.01$ dex.
These relatively small systematic errors do not affect
our results and discussion presented below.

\begin{figure*}[t]
\begin{center}
\includegraphics[width=18.5cm]{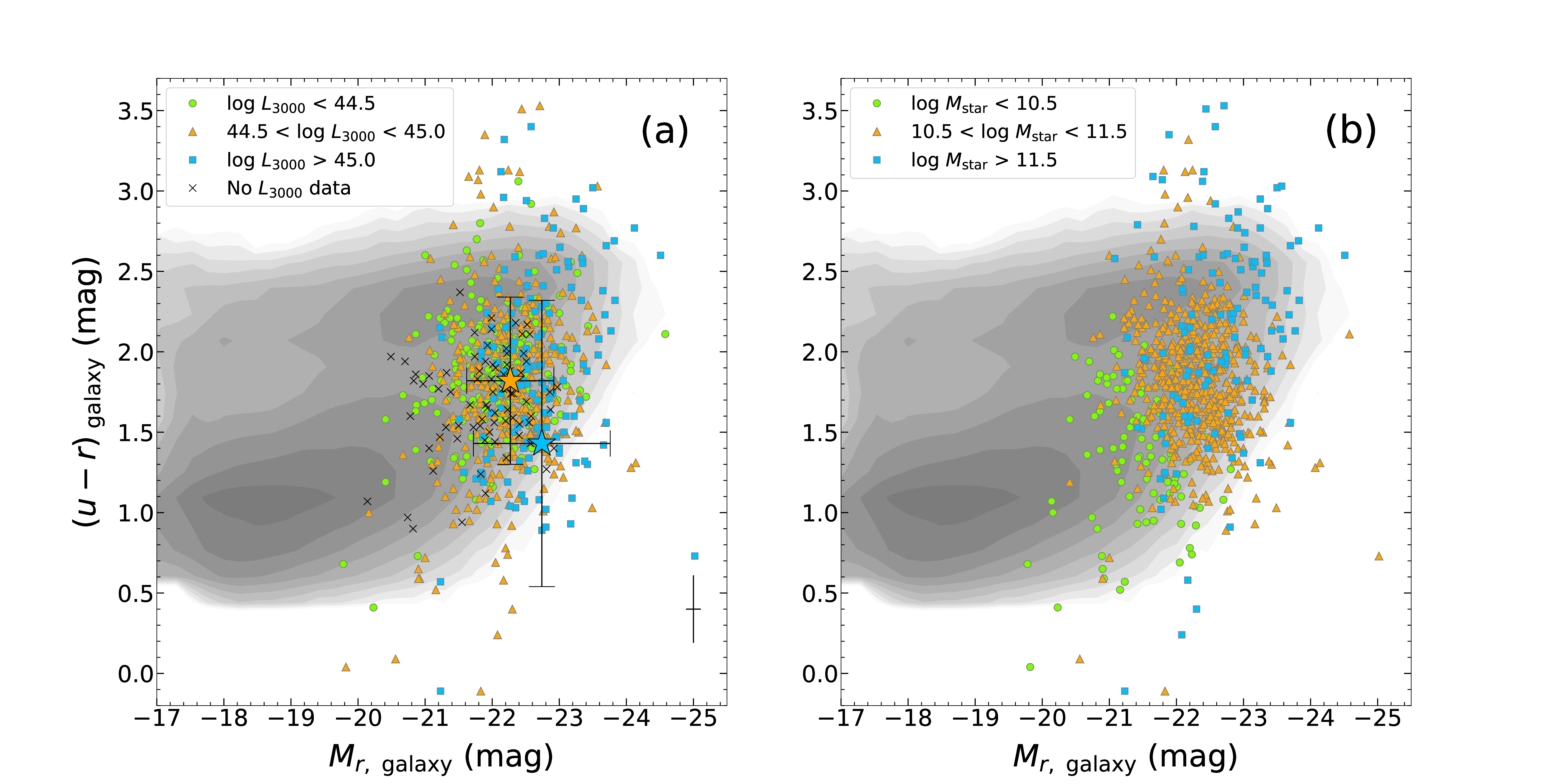} 
\end{center}
\caption{
(a) Rest-frame CMD of the quasar host galaxies, 
divided by nuclear continuum luminosities $L_{3000}$
(green circles; $\log{(L_{3000}/\mathrm{erg\ s^{-1}})}<44.5$, 
orange triangles; $44.5<\log{(L_{3000}/\mathrm{erg\ s^{-1}})}<45.0$, 
blue squares; $45.0<\log{(L_{3000}/\mathrm{erg\ s^{-1}})}$, 
crosses; no $L_{3000}$ data). 
A typical error bar is shown at the bottom right corner. 
The contours represent the distribution of non-AGN galaxies at $z<1$, 
selected from the HSC $\tt{Mizuki}$ photo-$z$ catalog (see text).
The orange and blue stars represent the median values of the main quasars
(mean $L_{3000} = 10^{44.7}\ \mathrm{erg\ s^{-1}}$) and saturated quasars (mean $L_{3000} = 10^{45.2}\ \mathrm{erg\ s^{-1}}$), respectively,
with the error bars representing the standard deviation of the sample scatter. 
(b) the CMD divided by $M_{\mathrm{star}}$
(green circles; $\log(M_{\mathrm{star}}/M_{\odot}) < 10.5$, 
orange triangles; $10.5 < \log(M_{\mathrm{star}}/M_{\odot}) < 11.5$, 
blue squares; $11.5 < \log(M_{\mathrm{star}}/M_{\odot})$).
}
\label{cmd}
\end{figure*}

\section{Results and Discussion}

\subsection{Stellar properties of the quasar host galaxies \label{subsec:cmd}}

Figure \ref{cmd} shows the rest-frame CMD of the quasar host galaxies, 
compared with the distribution of non-AGN galaxies at $z<1$ 
selected from the HSC $\tt{Mizuki}$ photometric redshift ($z_{\mathrm{photo}}$) catalog \citep{Tanaka15, Tanaka18}.
Here we regarded the objects with extended structures
($\tt{HSC\ classification\_extendedness=1}$) 
and high probabilities of objects being galaxy ($\tt{HSC\ prob\_gal}=1$),
as judged from broad-band SEDs
\footnote{The relative probabilities of objects being stars, galaxies, or quasars are derived by \cite{Tanaka18}
based on the SED-fitting using a set of template model SEDs.}
, as non-AGN galaxies;
hence the sample actually contains type-2 AGNs. 
We excluded those
objects with 68 \% confidence interval of $z_{\mathrm{photo}}$
larger than 0.2,
or the reduced $\chi^{2}$ of the SED-fitting larger than 5.0.
The systematic offset $(z_{\mathrm{photo}} - z_{\mathrm{spec}}) / (1+z_{\mathrm{spec}})$
and dispersion of the $\tt{Mizuki}$ $z_{\mathrm{photo}}$, 
with respect to spectroscopic redshifts, have been estimated to be 0.003 (0.013) and 0.048 (0.077), respectively,
in the VVDS-Deep (VVDS-UltraDeep) \citep{LeFevre13} fields \citep{Tanaka15}.
The non-AGN galaxies were selected in such a way that their redshift distribution matches that of the main sample quasars.

We found that the quasar host galaxies all have high luminosities
($M_{r,\mathrm{galaxy}}<-21$ mag), with respect to the non-AGN galaxies. 
This may in part be caused by our sample selection, 
which rejects $\sim25\%$ of the original HSC-SDSS quasars 
by the host-to-nuclear flux ratios (see Section \ref{subsec:image}). 
Alternatively, luminous quasars may intrinsically reside in more massive
and thus more luminous galaxies, 
which may in part reflect the $M_{\mathrm{BH}}-M_{\mathrm{star}}$ relation. 

The quasar hosts also show colors intermediate
between the blue cloud and the red sequence, 
i.e., they are mostly located in the green valley, 
which is believed to represent the transition phase 
from the blue cloud to the red sequence \citep{Salim14}.
This result is consistent with those of previous studies on
type-1 quasars \citep{Sanchez04, Matsuoka15}.
Considering the short timescale
that galaxies spend in the green valley
\citep[$<1$ Gyr;][]{Salim14},
that is evident from the sparseness of galaxies there,
the apparent clustering of quasar hosts implies that the quasar activity may make these galaxies green.
On the other hand, 
the process called ``rejuvenation", in which galaxies in the red sequence migrate to the green valley due to re-ignited 
star formation and AGN activities \citep[see, e.g.,][]{Suh19}, can also explain our results.
We found no apparent relation between the nuclear power ($L_{3000}$) 
and the location of the host galaxies on the CMD (Figure \ref{cmd} (a)); 
we get the same results when $L_{3000}$ is replaced
with $L_{[\mathrm{OI\hspace{-.1em}I\hspace{-.1em}I]}}$.
We also found that more massive hosts tend to have redder colors (Figure \ref{cmd} (b)),
which is consistent with the trend seen in non-AGN galaxies.
Note that there are some galaxies with extreme colors ($(u-r)_{\mathrm{galaxy}} < 0.5$
or $(u-r)_{\mathrm{galaxy}} > 3.0$), which are likely due to noise,
since these galaxies have rather high quasar-to-host contrast 
and thus have very large errors (typically $\sim2$ mag) in the derived rest-frame colors.

We confirmed that the host galaxies of
the saturated quasars have
an indistinguishable distribution
from the main sample on the CMD (see Figure \ref{cmd} (a)). 
The host galaxies of the saturated quasars have
the median magnitudes and colors 
($\langle M_{r,\mathrm{galaxy}} \rangle =-22.74 \pm 1.02$
and $\langle (u-r)_{\mathrm{galaxy}} \rangle = 1.43 \pm 0.89$)
consistent within 1$\sigma$ uncertainty with
those of the main sample
($\langle M_{r,\mathrm{galaxy}} \rangle =-22.27 \pm 0.65$
and $\langle (u-r)_{\mathrm{galaxy}} \rangle = 1.82 \pm 0.51$).
For reference, the median magnitude
and color of the merged main and saturated samples 
($\langle M_{r,\mathrm{galaxy}} \rangle =-22.30 \pm 0.72$ and 
$\langle (u-r)_{\mathrm{galaxy}} \rangle = 1.78 \pm 0.60$) are
almost identical to those of the main sample.
We conclude that excluding saturated quasars from the main sample does not change our results.

\citet{Kauffmann03} investigated the host galaxies of SDSS type-2 AGNs at $z < 0.3$.
They found that the hosts of AGNs with 
$L_{[\mathrm{OI\hspace{-.1em}I\hspace{-.1em}I]}} \gtrsim 10^{40}\ \mathrm{erg\ s^{-1}}$
are located in the middle position 
between star-forming galaxies and passive galaxies  
on the 4000 \AA\ break 
(a measure of stellar population age)
versus $M_{\mathrm{star}}$ plane.
We found that our quasar hosts, when limited to
$L_{[\mathrm{OI\hspace{-.1em}I\hspace{-.1em}I]}} \gtrsim 10^{40}\ \mathrm{erg\ s^{-1}}$,
are also mainly located on the green valley.
The hosts of the \cite{Kauffmann03} sample are widely spread from the blue cloud to the red sequence,
but the scatter looks smaller than in our sample; 
this is presumably due to the uncertainty introduced by our decomposition analysis,
as the type-2 AGNs allow direct measurements of the host galaxies
without such a decomposition procedure.

\citet{Goulding14} investigated the host galaxies of IR and X-ray AGNs,
with similar bolometric luminosities ($\sim 10^{44.5-46.5}\ \mathrm{erg/s}$)
to our main sample
\footnote{The bolometric luminosity of the \citet{Goulding14} sample was estimated from 
the 4.5 $\mu \mathrm{m}$ or X-ray (0.5-7 keV) luminosity, following the method in \citet{Hickox09}.}.
Their host galaxies tend to 
reside in the blue cloud to the green valley,
peaking at the green valley,
which is consistent with the present results.
The scatter in color is larger in our sample,
which needs the decomposition analysis to measure the hosts
(while the obscured AGNs in \citet{Goulding14} do not).
They pointed out that
galaxies in the green valley may be affected by dust reddening.
Indeed, the $\tt{CIGALE}$ best-fit SED models of our sample indicate that 
$\sim55\%$ of the host galaxies have $A_{V}>1$ mag, 
and that the host galaxies have blue colors similar to star-forming galaxies 
when corrected for dust reddening.
However, the present estimates of $A_{V}$ values are not very accurate,
given that they are based on only five broad-band magnitudes.
\cite{Pacifici12} reported that the $A_{V}$ estimates based on optical broad-band photometry have 
typical systematic and random uncertainty of 0.23 and 1.0 mag, respectively.

\citet{Suh19} investigated type 1 and 2 AGNs in the $Chandra$-COSMOS Legacy Survey. 
They performed quasar/galaxy decomposition with SED fitting of the multi-band photometry data, 
and derived SFRs from far-IR luminosities using $\it{Spitzer}$ and $\it{Herschel}$ data. 
They found that the host galaxies of the both types lie on the star-forming main sequence. 
Their result indicates that AGNs do not supress the star formation activities of their host galaxies, 
and seems inconsistent with our result.
The present study focuses on optical wavelengths, 
and cannot trace the IR dust emission from star formation. 
Moreover, we may have subtracted an unresolved nuclear starburst region
 with the PSF component in the image decomposition, 
that would make the colors of the host galaxies redder than actually are. 
On the other hand, \citet{Symeonidis16} claimed 
that far-IR emission includes a significant contribution from quasar nucleus, 
and thus is not a good tracer of star formation. 
If this is true, then \citet{Suh19} might have overestimated SFRs of the host galaxies.

\begin{figure}	
\begin{center}
\includegraphics[width=9.1cm]{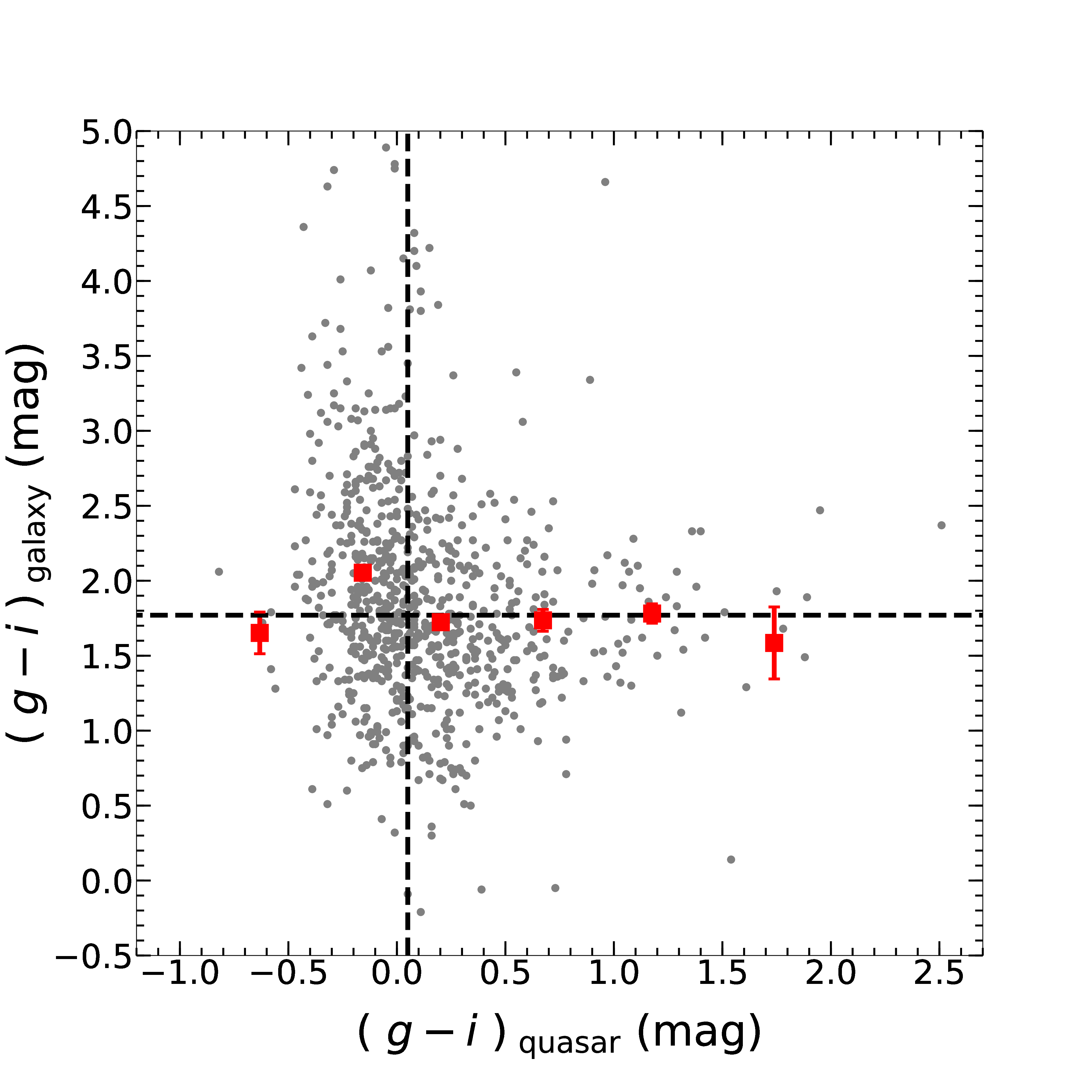} 
\end{center}
\caption{
The observed $g-i$ colors of the host galaxies ($\mathrm{S\acute{e}rsic}$ component)
against those of the quasar nuclei (PSF component). 
The squares represent the mean values of the host galaxy colors
in bins of nuclear color. 
The vertical and horizontal dashed lines represent the median colors of the nuclei
and the host galaxies, respectively.
Note that the host galaxies with $(g-i)_{\mathrm{galaxy}} > 3.0$
have large errors in the color measurements,
and thus their extreme colors are likely due to noise. }
\label{color}
\end{figure}

Figure \ref{color} presents the observed $g-i$ colors of the host galaxy components 
against those of the nuclear components. 
There is little correlation, 
which suggests that the nuclear contamination to the host flux 
due to imperfect image decomposition, 
or the effect of scattered nuclear light in the hosts, is not significant.
In order to further test the possibility of such contamination, 
we include the \citet{Fritz06} AGN model spectrum
in the SED-fitting of the host galaxy fluxes.
The mean contributions of AGN in the decomposed host galaxy fluxes are $15\%$ and $6\%$ 
in the rest-frame $u$ and $r$-bands, respectively.
About 70\% of the host galaxies are best fitted without AGN contribution.
We found that 
the host galaxies are consistently found in the green valley
even if the SED fitting includes the AGN template,
suggesting that AGN contribution dose not significantly affect our conclusions.

\subsection{The quasar to galaxy flux ratio}

\begin{figure}	
\begin{center}
\includegraphics[width=8.5cm]{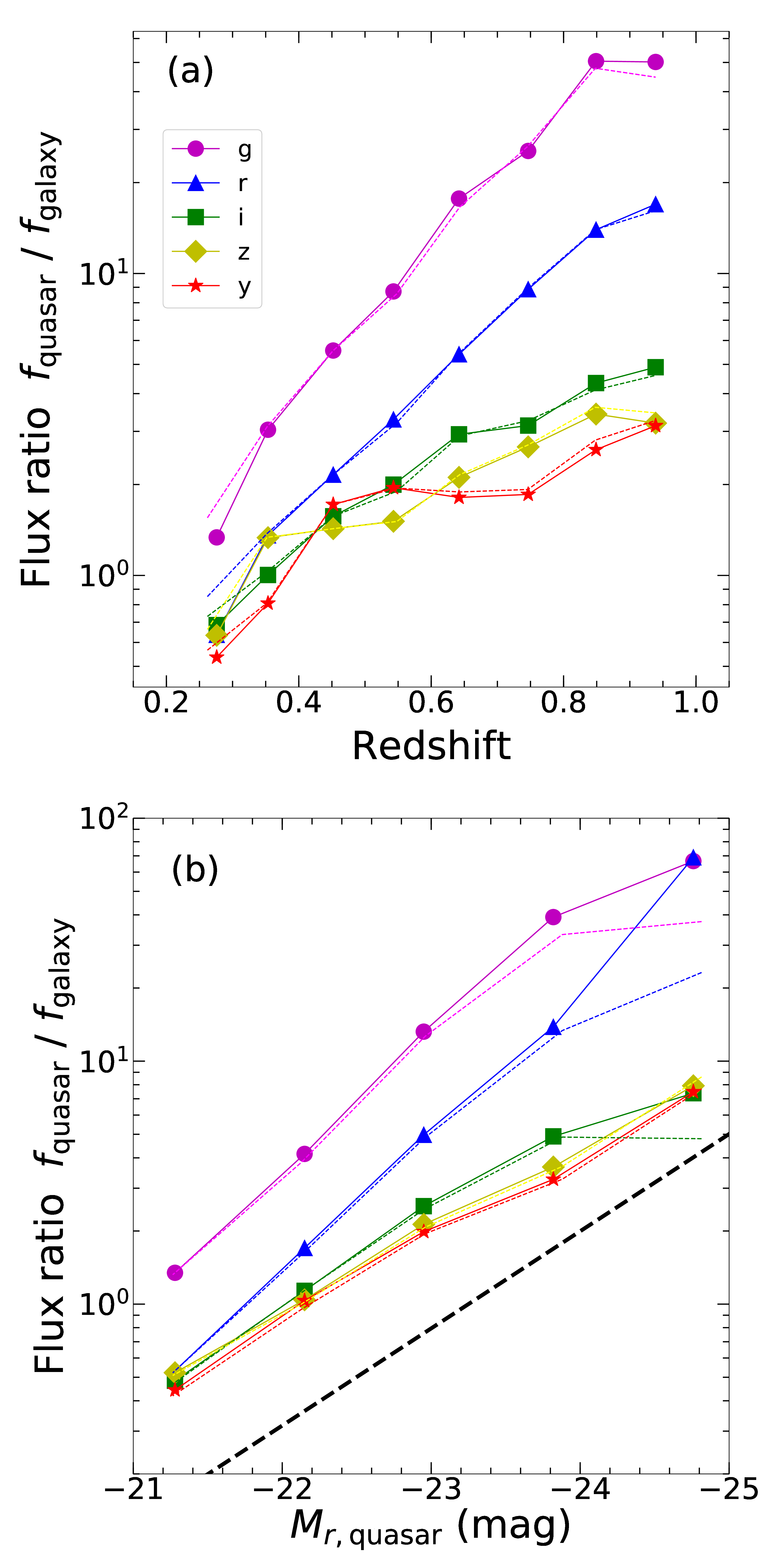} 
\end{center}
\caption{
(a) The median quasar-to-host-galaxy flux ratios ($f_{\mathrm{quasar}}/f_{\mathrm{galaxy}}$) in the five bands
($g$; pink circles, $r$; blue triangles, $i$; green squares, $z$;  yellow diamonds, $y$; red stars), as a function of redshift.
The symbols and solid lines represent the main sample,
while the dotted lines represent the merged main and saturated samples.
(b) $f_{\mathrm{quasar}}/f_{\mathrm{galaxy}}$ as a function of 
the $r$-band absolute magnitudes of the quasar components.
The symbols are same as (a).
The dashed line represents the slope of constant host galaxy flux.
}
\label{flux_ratio}
\end{figure}

Figure \ref{flux_ratio} (a) shows the median values of the quasar-to-host-galaxy flux ratios, 
$f_{\mathrm{quasar}}/f_{\mathrm{galaxy}}$,
as a function of redshift in the five bands. 
As expected, the flux ratios are larger in the bluer bands.
This trend remains unchanged when the saturated quasars are included.
The correlation between redshifts and the flux ratios is due to a selection bias;
at a fixed host galaxy flux, fainter quasars are more difficult to detect at higher redshifts. 
Furthermore, a given band traces a bluer part of the spectrum for high-$z$ quasars,
and $f_{\mathrm{quasar}}/f_{\mathrm{galaxy}}$ is usually larger
at shorter wavelengths \citep[e.g.,][]{Matsuoka14}.

Figure \ref{flux_ratio} (b) presents the flux ratios 
as a function of $M_{r,\mathrm{quasar}}$. 
The plotted curves of the main sample are almost parallel
to the line of constant host galaxy flux, 
which indicates that 
the host galaxy luminosities are independent of the nuclear luminosities.
On the other hand, we will see below that
$M_{r,\mathrm{galaxy}}$ and $M_{\mathrm{BH}}$ are correlated
(Section 4.3).
If quasars with a given $M_{\mathrm{BH}}$ radiate 
in a narrow range of Eddington ratio,
the nuclear luminosity and $M_{\mathrm{BH}}$ should correlate.
Then, the nuclear and the host luminosity should correlate,
considering the $M_{\mathrm{BH}}-M_{\mathrm{star}}$ relation.
Therefore, the independence of the nuclear and host luminosities indicates that 
quasars radiate over a wide range of Eddington ratio.
The independence of nuclear and host galaxy luminosities is also reported 
for Seyfert galaxies at $z < 0.15$ \citep{Hao05} 
and for (type-1) quasars at $z<0.5 - 0.6$ \citep{Falomo14, Matsuoka14}.
The flux ratios are flatter at $M_{r,\mathrm{quasar} }< -24$ mag
when the saturated quasar sample is included.
This may indicate that the host galaxies become relatively more luminous with increasing nuclear luminosities,
in the most luminous quasars, 
but we have no plausible explanation for this trend if it is real.

Note that we have excluded from the analysis 
the quasars with large $f_{\mathrm{quasar}}/f_{\mathrm{galaxy}}$,
whose host galaxies are hard to decompose.
Therefore, the intrinsic slopes may be steeper than 
what appear in Figure \ref{flux_ratio} (b).
Also, our sample does not include quasars less luminous than 
$M_{r,\mathrm{quasar}} \simeq -21$ mag, which have been excluded by the selection criterion used in the SDSS quasar catalog.
Thus the slopes at $M_{r,\mathrm{quasar}} < -21$ mag may be different from what we see at the brighter side in Figure \ref{flux_ratio} (b),
but that issue is beyond the scope of this paper.

\subsection{The $M_{\mathrm{BH}}-M_{\mathrm{star}}$ relation}

\begin{figure}	
\begin{center}
\includegraphics[width=9.0cm]{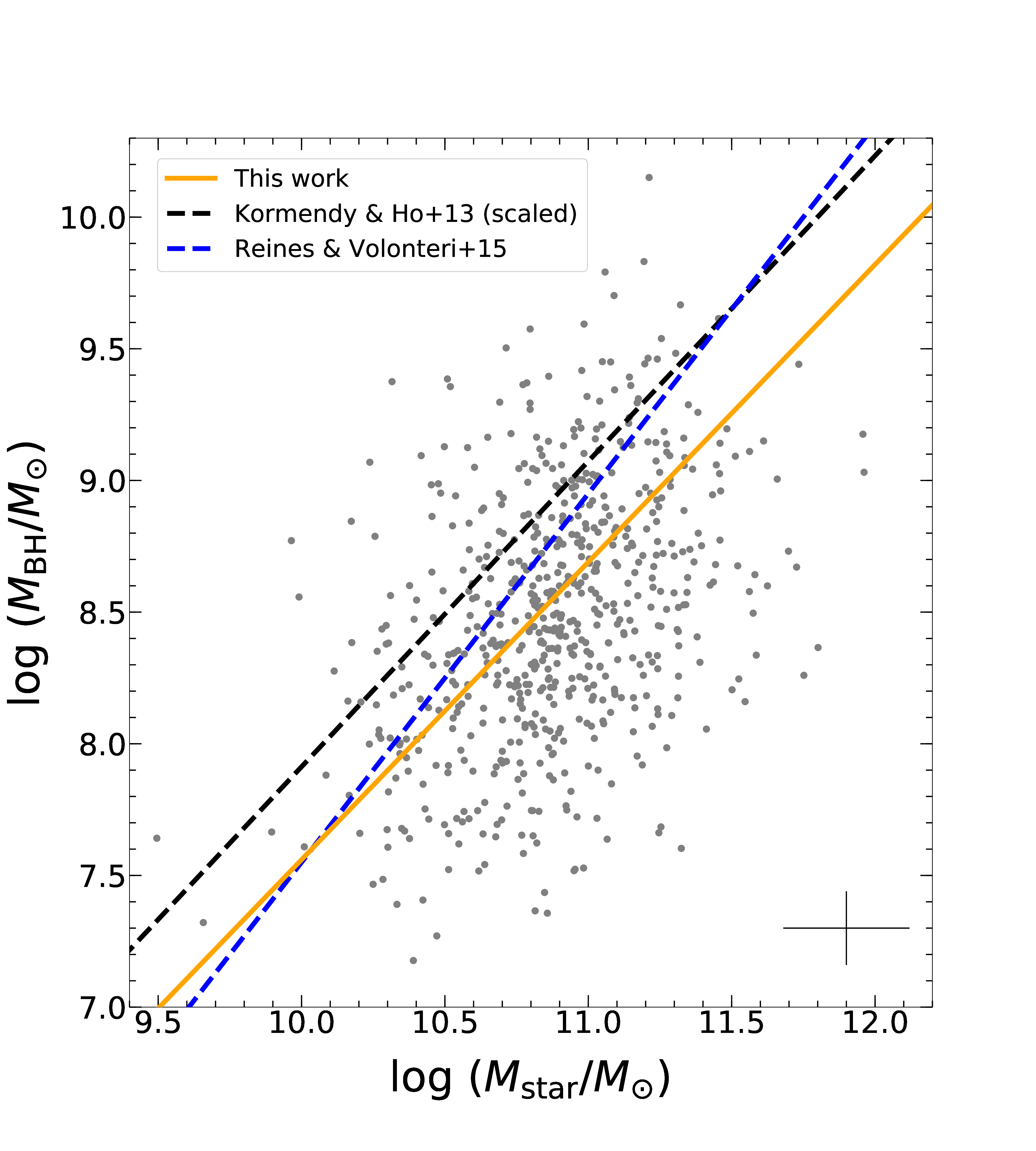} 
\end{center}
\caption{
Relation between $M_{\mathrm{star}}$ and $M_{\mathrm{BH}}$ of
the 646 quasar host galaxies with $\Delta \log(M_{\mathrm{star}}/M_{\odot}) < 0.5$.
The typical error is shown at the bottom right.
The best-fit line is represented by the orange solid line. 
The black line represents the local $M_{\mathrm{BH}}-M_{\mathrm{bulge}}$ relation of \citet{KH13},
and $M_{\mathrm{bulge}}$ of this relation is scaled down by 0.33 dex
to take into account the deference of the IMF between our study and \citet{KH13} (see the text).
The blue dashed line represents the local  $M_{\mathrm{BH}}-M_{\mathrm{star}}$ relation of \citet{RV15}.
}
\label{Mstar_Mbh}
\end{figure}

Figure \ref{Mstar_Mbh} shows the relation
between $M_{\mathrm{star}}$ and $M_{\mathrm{BH}}$
of our main quasar sample, excluding those with large errors in
the $\tt{CIGALE}$ stellar mass estimates, 
$\Delta \log(M_{\mathrm{star}}/M_{\odot}) > 0.5$.
For comparison, we plot the local $M_{\mathrm{BH}}-M_{\mathrm{bulge}}$ relation 
presented in \citet{KH13}.
We shifted $M_{\mathrm{bulge}}$ of \citet{KH13} by $-$0.33 dex
according to \citet{RV15}, in order to take into account the difference of the adopted IMFs between \citet{KH13}
\citep{Kroupa01} and the present study \citep{Chabrier03}.
We note that the $M_{\mathrm{star}}$ values in our sample includes 
both bulge and disk components.
So if a disk is present, $M_{\mathrm{star}}$ represents 
an upper limit to $M_{\mathrm{bulge}}$. 
We also plot the local $M_{\mathrm{BH}}-M_{\mathrm{star}}$ relation of \citet{RV15}, 
for a sample of elliptical galaxies and galaxies with classical bulges.
We derived the regression line of our data points
using a Bayesian maximum likelihood method of \citet{Kelly07},
which was also used by \citet{RV15}.
This method takes into account the errors
in both independent ($M_{\mathrm{star}}$) 
and dependent ($M_{\mathrm{BH}}$) variables,
and also intrinsic scatter around the regression line.
It assumes uniform prior distributions for the parameters of regression line,
and performs the Markov chain Monte Carlo simulation.
We use median values and standard deviations
taken from the posterior probability distribution 
as the best-fit line parameters and uncertainties, respectively.
The best-fit regression line is
	\begin{eqnarray}
		\log{ \Big(\frac{M_{\mathrm{BH}}}{10^{8}M_{\odot}} \Big) } &=& 
		(1.13 \pm 0.14)\ \log{ \Big( \frac{M_{\mathrm{star}}}{10^{11}M_{\odot}} \Big) } + (0.69 \pm 0.03),
		\label{eq:mbh_mstar}
	\end{eqnarray}
with the intrinsic scatter of 0.35 dex.
We also derived the regression line 
with another algorithm,
the symmetrical least-square method of \citet{Tremaine02},
which was used by \citet{KH13}. 
We confirmed that the slope and intercept remain unchanged
with this alternative fitting method.

Our $M_{\mathrm{BH}}-M_{\mathrm{star}}$ relation has a similar slope with the local relations,
while $M_{\mathrm{BH}}$ at a given $M_{\mathrm{star}}$ seems lower than those in the local universe.
Several biases affect this result.
For a given $M_{\mathrm{BH}}$, less luminous/massive hosts tend to be excluded from the sample
by our selection criteria (Section \ref{subsec:image}).
Furthermore, \citet{Schulze15} found that
the active fraction (the AGN fraction among the whole SMBHs)
decreases with increasing $M_{\mathrm{BH}}$ at $z<1$.
This would bias against massive SMBH in a given AGN sample, at a given $M_{\mathrm{star}}$.
Only 56 \% of the initial 1151 objects were used for the present fitting of the $M_{\mathrm{BH}}-M_{\mathrm{star}}$ relation,
while no systematic difference was found in the $M_{\mathrm{BH}}$ distribution between the included and excluded objects.
We cannot rule out the possibility that those excluded quasars may have different $M_{\mathrm{BH}}-M_{\mathrm{star}}$ relation,
which can be tested only with higher quality data.
 
\section{Summary}

We studied the properties of 862 quasar host galaxies at $z<1$, 
using the SDSS quasar catalog
and the Subaru HSC images in five bands ($g,r,i,z,$ and $y$)
with a typical PSF size of $0.6$ arcsec.
The unprecedented combination of the survey area and depth allows us to 
perform a statistical analysis of the quasar host galaxies, with small sample variance.
We decomposed the quasar images into the nuclei and the host galaxies,
using the PSF and the $\mathrm{S\acute{e}rsic}$ models. 
The host components are detected in more than three bands 
in almost all the cases, 
which allowed us to perform SED fitting of the host galaxy fluxes. 
Our main results are as follows:
\begin{enumerate}

	\item The quasar host galaxies are mostly luminous ($M_{r}<-21$ mag) and are located in the green valley. 
	This is consistent with a picture in which AGN activity suppresses star formation of the host galaxies (i.e., AGN feedback), 
	and drives the migration of galaxies from the blue cloud to the red sequence.
	
	\item The host galaxy luminosities have little dependence on the nuclear luminosities.
	
	\item The $M_{\mathrm{BH}}-M_{\mathrm{star}}$ relation of the quasars has a consistent slope with the local relations,
	while the SMBHs may be slightly undermassive.
	However, this result is subject to our sample selection, which biases against host galaxies with low masses and/or
	large quasar-to-host flux ratios.
	
\end{enumerate}

The HSC-SSP survey is still ongoing, and 
the wider sky coverage will allow us to further expand the sample size.
We also plan to investigate quasar host galaxies
in the HSC-SSP Deep/UltraDeep data,
which are deeper by about $1\sim$2 mag than 
the Wide data used in this work.
The deeper data will allow for more accurate decomposition of quasar images, including the sample
with faint host galaxies that we had to exclude in the present analysis.

\begin{ack}
We are grateful to the referee for his/her useful comments to improve this paper.

Y.M. was supported by the Japan Society for the Promotion of Science (JSPS) KAKENHI grant No. JP17H04830 and the Mitsubishi Foundation grant No. 30140.

The Hyper Suprime-Cam (HSC) collaboration includes the astronomical communities of Japan and Taiwan, and Princeton University. 
The HSC instrumentation and software were developed by NAOJ, 
the Kavli Institute for the Physics and Mathematics of the Universe (Kavli IPMU), 
the University of Tokyo, the High Energy Accelerator Research Organization (KEK), 
the Academia Sinica Institute for Astronomy and Astrophysics in Taiwan (ASIAA), and Princeton University. 
Funding was contributed by the FIRST program from Japanese Cabinet Office, the Ministry of Education, Culture, Sports, Science and Technology (MEXT), 
the Japan Society for the Promotion of Science (JSPS), Japan Science and Technology Agency (JST), 
the Toray Science Foundation, NAOJ, Kavli IPMU, KEK, ASIAA, and Princeton University.

This paper makes use of software developed for the Large Synoptic Survey Telescope. 
We thank the LSST Project for making their code available as free software at http://dm.lsstcorp.org.

The Pan-STARRS1 Surveys (PS1) have been made possible through contributions of the Institute for Astronomy, 
the University of Hawaii, the Pan-STARRS Project Office, 
the Max-Planck Society and its participating institutes, the Max Planck Institute for Astronomy, 
Heidelberg and the Max Planck Institute for Extraterrestrial Physics, Garching, 
The Johns Hopkins University, Durham University, the University of Edinburgh, 
Queen's University Belfast, the Harvard-Smithsonian Center for Astrophysics, 
the Las Cumbres Observatory Global Telescope Network Incorporated, the National Central University of Taiwan, 
the Space Telescope Science Institute, the National Aeronautics and Space Administration under Grant No. 
NNX08AR22G issued through the Planetary Science Division of the NASA Science Mission Directorate, 
the National Science Foundation under Grant No. AST-1238877, the University of Maryland, and Eotvos Lorand University (ELTE).

\end{ack}

%


\end{document}